\documentclass[12pt,preprint]{aastex}
\shorttitle{ASAS J002511+1217.2: A New WZ Sagittae Star?}
\shortauthors{Templeton et al.}

\begin{document}
\title{The Recently-Discovered Dwarf Nova System ASAS J002511+1217.2: A New
WZ Sagittae Star\footnote{Based in part on observations obtained with the
Apache Point Observatory (APO) 3.5m telescope, which is owned and operated by
the Astrophysical Research Consortium (ARC).}}
\author{M.R. Templeton\altaffilmark{2}, 
R. Leaman\altaffilmark{3},
P. Szkody\altaffilmark{3},
A. Henden\altaffilmark{2,4},
L. Cook\altaffilmark{5},
D. Starkey\altaffilmark{2},
A. Oksanen\altaffilmark{6},
M. Koppelman\altaffilmark{2},
D. Boyd\altaffilmark{7},
P.R. Nelson\altaffilmark{8}
T. Vanmunster\altaffilmark{9},
R. Pickard\altaffilmark{10},
N. Quinn\altaffilmark{11},
R. Huziak\altaffilmark{2},
M. Aho\altaffilmark{6},
R. James\altaffilmark{2},
A. Golovin\altaffilmark{12},
E. Pavlenko\altaffilmark{13},
R.I. Durkee\altaffilmark{14},
T.R. Crawford\altaffilmark{2,15},
G. Walker\altaffilmark{2}, \&
P. P\"{a}\"{a}kk\"{o}nen\altaffilmark{16}
}
\altaffiltext{2}{AAVSO, 25 Birch Street, Cambridge, MA 02138}
\altaffiltext{3}{Department of Astronomy, University of Washington, Box 351580, Seattle, WA 98195}
\altaffiltext{4}{Universities Space Research Association/U.S. Naval Observatory,
P.O. Box 1149, Flagstaff, AZ 86002-1149}
\altaffiltext{5}{Center for Backyard Astrophysics -- Concord, 1730 Helix Ct.,
Concord, CA 94518}
\altaffiltext{6}{Nyr\"{o}l\"{a} Observatory, Kyllikinkatu 1, FI-40100 Jyvaskyla,
Finland}
\altaffiltext{7}{British Astronomical Association Variable Star Section, West
Challow, OX12 9TX, United Kingdom}
\altaffiltext{8}{RMB 2493, Ellinbank, Victoria 3820, Australia}
\altaffiltext{9}{CBA Belgium Observatory, Walhostraat 1A, B-3401 Landen, Belgium}
\altaffiltext{10}{British Astronomical Association Variable Star Section,
3 The Birches, Shobdon, Leominster, Herefordshire HR6 9NG England}
\altaffiltext{11}{British Astronomical Association Variable Star Section,
Steyning, West Sussex, BN44 3LR, England}
\altaffiltext{12}{Kyiv National Shevchenko University, Physics Department;
Visiting Astronomer, Crimean Astrophysical Observatory, Komunarov 63, {\#}39,
Berdyansk, Zaporozhskaja, 71118, Ukraine}
\altaffiltext{13}{Crimean Astrophysical Observatory, Nauchny, Crimea, 98409 Ukraine}
\altaffiltext{14}{Shed of Science Observatory, 5213 Washburn Ave S., 
Minneapolis, MN 55410}
\altaffiltext{15}{Arch Cape Observatory, Arch Cape, OR}
\altaffiltext{16}{Amateur Astronomer Association Seulaset r.y., University of
Joensuu, P.O. Box 111, FI-80101 Joensuu, Finland}

\begin{abstract}
The cataclysmic variable ASAS J002511+1217.2 was discovered in outburst by the
All-Sky Automated Survey in September 2004, and intensively monitored by AAVSO
observers through the following two months.  Both photometry and spectroscopy
indicate that this is a very short-period system.  Clearly defined superhumps 
with a period of $0.05687 \pm 0.00001$ (1-$\sigma$) days (81.9 minutes) are 
present during the superoutburst, 5 to 18 days following the ASAS detection.
We observe a change in superhump profile
similar to the transition to ``late superhumps'' observed in 
other short-period systems; the superhump period appears to increase slightly
for a time before returning to the original value, with the resulting superhump
phase offset by approximately half a period.  We detect variations with a 
period of $0.05666 \pm 0.00003$ (1-$\sigma$) days (81.6 minutes) during the 
four-day quiescent phase between the end of the main outburst and the single 
echo outburst.  Weak variations having the original superhump period reappear
during the echo and its rapid decline.  Time-resolved spectroscopy conducted 
nearly 30 days after detection and well into the decline yields an orbital 
period measurement of $82 \pm 5$ minutes.  Both narrow and broad components 
are present in the emission line spectra, indicating the presence of multiple 
emission regions.  The weight of the observational evidence suggests that 
ASAS J002511+1217.2 is a WZ Sge-type dwarf nova, and we discuss how this 
system fits into the WZ classification scheme.
\end{abstract}


\section{Introduction}

On 2004 September 11, the All-Sky Automated Survey \citep{poj02} detected the
transient object ASAS J002511+1217.2 ($\alpha$: 00h 25m 11.4s, $\delta$ 
+12$^{\circ}$ 17' 14.3", J2000), a source not previously known to be variable 
(see \citet{price04} for details).  It was later found that the object at this 
position had previously been detected by ROSAT (1RXS J002510.8+121725; 
\citet{voges99}) and had been identified as a blue point source in the 
Hamburg Quasar Survey \citep{zick03}.  Amateur and professional astronomers 
from around the world began observing this object, and tracked its behavior for
two months following the initial detection.  
Time-series spectra of ASAS J002511+1217.2 were obtained late in the outburst
on JD 2,453,288 (2004 October 10) using the Apache Point Observatory 3.5-meter 
telescope.  The spectra, obtained over a span of two hours, show radial 
velocity variations consistent with an orbital period of approximately 82 
minutes.  The photometric and spectroscopic behavior of this object is 
clearly that of a dwarf nova system, having many of the hallmarks of the small 
class of WZ Sagittae stars \citep{golovin05}.  The WZ Sge stars are 
characterized by large-amplitude, infrequent outbursts, orbital periods below 
the period gap (around 2 hours), the presence of superhumps in the light curve, 
echo outbursts, and a slow decline from maximum light to quiescence.  For an 
overview of the WZ Sge class, see \citet{warner95}.

In this paper, we discuss the available photometric and spectroscopic evidence
and justify our classification of ASAS J002511+1217.2 as a WZ Sge star.  In 
Section 2, we describe the photometric observations obtained by AAVSO observers
during the 2004 outburst and discuss the periodic behavior seen.  In 
Section 3, we describe the spectroscopic observations, and the orbital 
parameters derived from these data.  In Section 4, we discuss how this star
fits into the WZ Sge class, and what questions remain about its nature.

\section{Photometry}

\begin{figure}
\begin{center}
\includegraphics[width=0.95\textwidth]{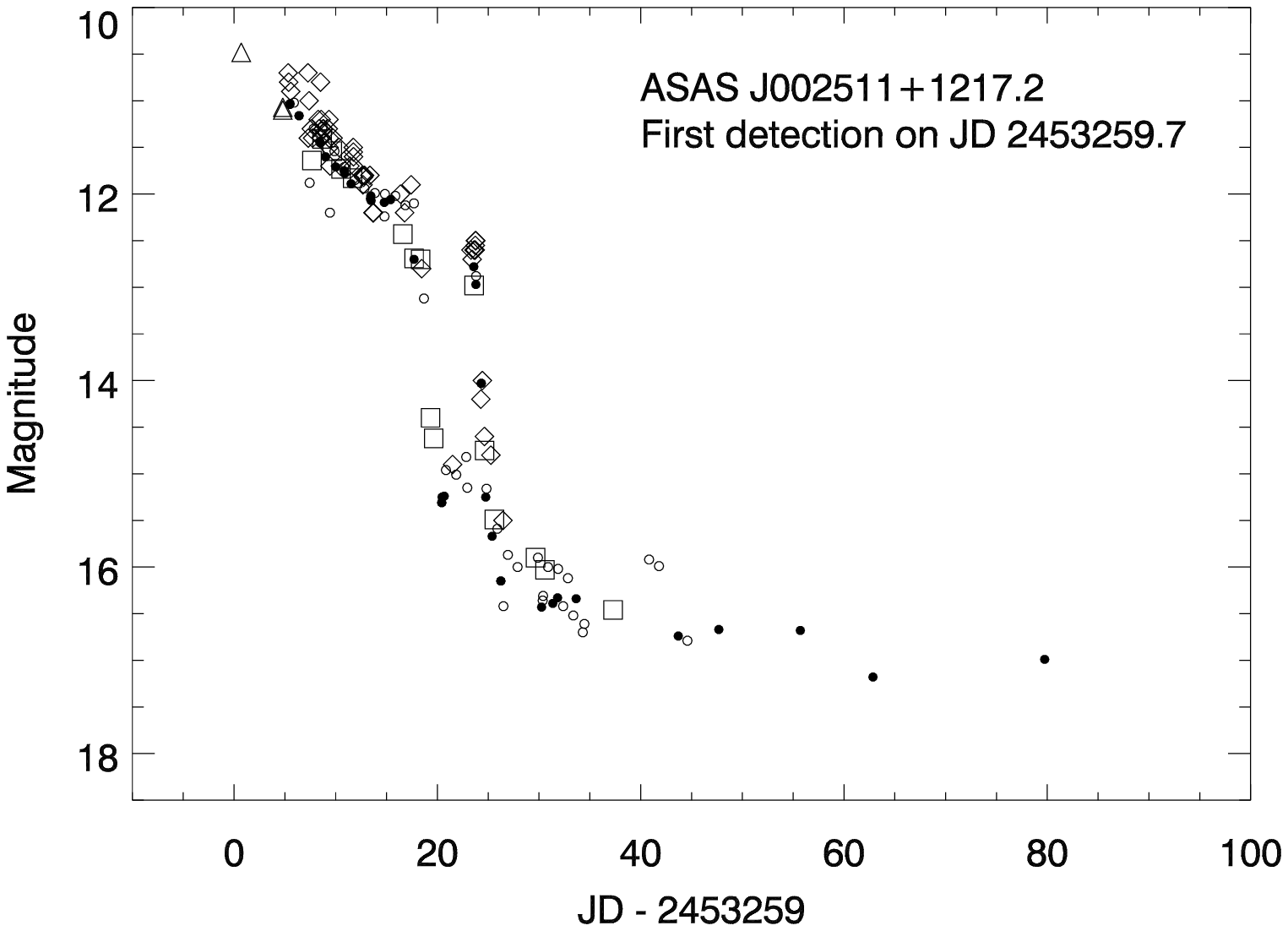}
\caption{Outburst light curve of ASAS J002511+1217.2.  The outburst amplitude
is nearly seven magnitudes from the presumed quiescent level of $V \sim 17.3$.
The main ``superoutburst'' lasted for at least 18 days, and one echo outburst 
was observed approximately 5 days after the initial outburst ended.  Following
the echo outburst, the star dropped below $V \sim 16$, and slowly faded to
$V \sim 17$ through the end of 2004.  The data points are nightly averages
of individual observers (CCD observations) or individual visual estimates.
Open diamonds: visual magnitude estimates; open circles: unfiltered CCD 
observations; filled circles: $V$-band CCD observations; open squares: 
$R$-band CCD observations; open triangles: {\it ASAS} $V$-band observations.
We note that the data are not on a common photometric system, and some of the
variations are due to observer-to-observer zero-point differences of a few
tenths of a magnitude.  This is particularly the case with unfiltered CCD 
observations, where the spectral response of individual observers systems 
can vary widely.  All linear trends and zero-point differences are subtracted 
from individual observers' nightly time-series data prior to analysis.}
\label{Fig01}
\end{center}
\end{figure}

The photometric observations of ASAS J002511+1217.2 were first discussed by
\citet{golovin05}, and were obtained during the course of the outburst 
beginning on 2004 September 10.  Amateur and professional observers from 
around the world contributed nearly 17000 
positive observations of ASAS J002511+1217.2 between 2004 September 15 
(JD 2,453,264) and 2004 December 12 (JD 2,453,352).  In addition, a few hundred 
visual and CCD ``fainter than'' estimates were obtained through 2005 
February 13 (JD 2,453,414.5). 
The majority of observations were made with CCD cameras, both filtered and 
unfiltered.  Figure \ref{Fig01} shows the overall
light curve of this variable, including all of the positive visual observations
and un-calibrated CCD measurements.  Of particular interest are the high 
amplitude of the initial superoutburst (nearly seven magnitudes), the early,
abrupt decline in the superoutburst at JD 2,453,277 followed by a single echo
outburst at JD 2,453,282, and the long decay time to quiescence (90 days).  The
high outburst amplitude, long decay time, and echo outbursts are phenomena
observed in the WZ Sge stars, and are our primary reasons for suggesting
the WZ Sge classification.  

In the following subsections, we discuss the photometric behavior both early 
and late in the outburst, concentrating on CCD observations made early in the 
outburst to detect and characterize the superhump period and evolution.  We 
also discuss the late-outburst photometry, the decline toward quiescence and
the probable quiescent level.

\subsection{Early outburst: superhumps}

\begin{figure}
\begin{center}
\includegraphics[width=0.95\textwidth]{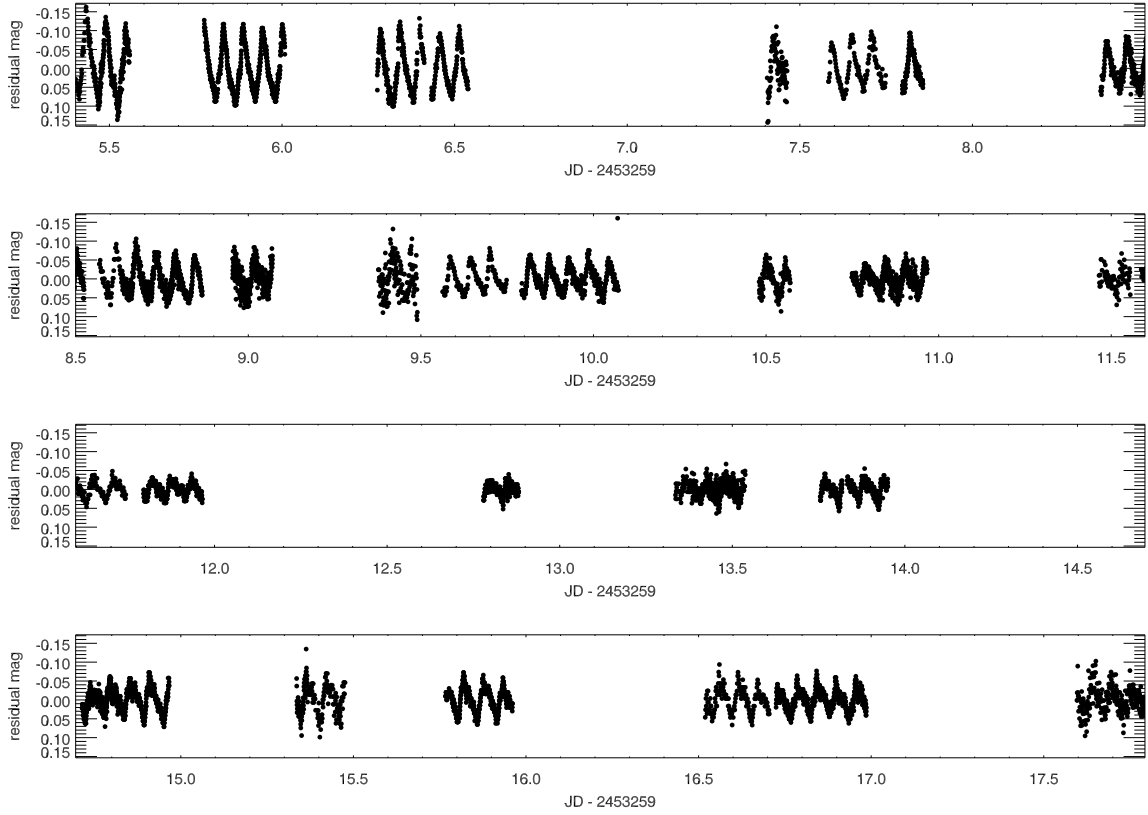}
\caption{Superhump light curve of ASAS J002511+1217.2 from JD 2,453,264 to
2,453,276.  The light curve shows significant evolution of the superhump 
profile and amplitude over the course of the outburst.}
\label{Fig02}
\end{center}
\end{figure}

The photometric data were obtained by many amateur and professional observers
world-wide, and were taken with a variety of equipment including telescopes of 
both large and small aperture, and with a variety of filters and CCD cameras.  
However, for period-searching the absolute calibration of the data is not
very important.  Therefore, prior to time-series analysis, a simple linear fit
was subtracted from each observer's data on each night to remove the much 
larger variation of the underlying outburst along with any zero point 
differences.  The detrended light curve of the superhumps is shown in Figure 
\ref{Fig02}.  As we discuss below, our time-series analysis indicates a
superhump period of 0.0569 days (81.9 minutes).  We do not detect any eclipse
signature in the lightcurve, and any signal caused by orbital variation
would be difficult to disentangle from the superhump and its evolution.  For
now, we treat these data as if the signal were purely due to superhumps.

\subsubsection{Fourier analysis}

Because the periods and light curve profiles of superhumps are known to
change throughout the course of outbursts, we investigated whether or not
period and superhump profile variations were apparent in the light curve of
ASAS J002511+1217.2.  We performed two separate Fourier analyses on the data: 
one in which we analyzed single days of photometry individually, and one in 
which we analyzed {\it clusters} of frequencies in the full data set's Fourier 
transform, to search for time-varying periods in the complete set of data.  
Signals with time-varying periods will generate clusters of frequencies in a
Fourier power spectrum, rather than a single peak at the
main frequency, and analysis of these clusters can provide information on the
period evolution with time.  See \citet{foster95} for a description of this
method.

\begin{figure}
\begin{center}
\includegraphics[width=0.95\textwidth]{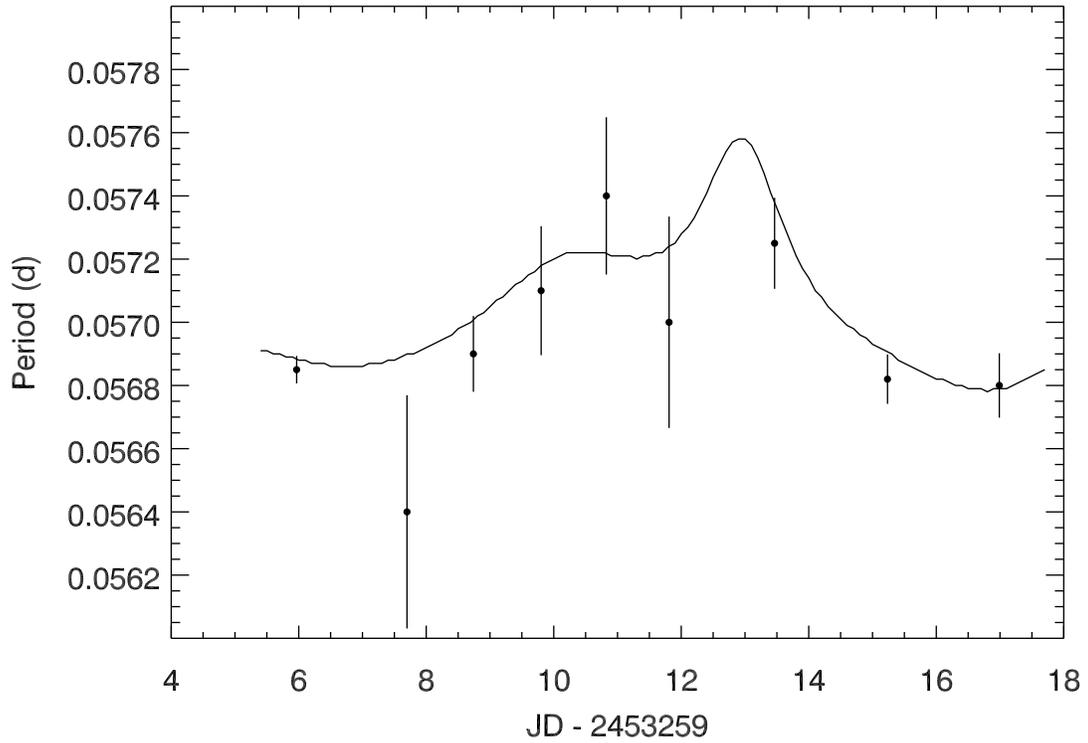}
\caption{Period change in the superhumps during the early outburst.
Points: periods derived from single days of photometry, with
two-sigma error bars; solid line: period derived using the cluster analysis
of \citet{foster95}.  The period of the superhumps undergoes statistically
significant period change beginning on the fifth day of photometry, 
approximately 8 days after the outburst was first detected.}
\label{Fig03}
\end{center}
\end{figure}

Figure \ref{Fig03} shows the period as a function of time for the two analysis
methods.  Because the photometric coverage of individual days was not uniform,
the error bars of the daily period measurements differ throughout.  However, 
the period clearly varies at the 3-$\sigma$ level.  Variations were strongly 
detected by the first reported time-series observations on JD 2,453,264, five 
days after the outburst was first detected.  Fourier analysis indicates an 
average period of 0.0569 days, which remained reasonably constant until JD 
2,453,268 (nine days after the outburst detection) when the period began to 
increase by a small but significant amount.

\subsubsection{(O-C) analysis}

\begin{figure}
\begin{center}
\includegraphics[width=0.95\textwidth]{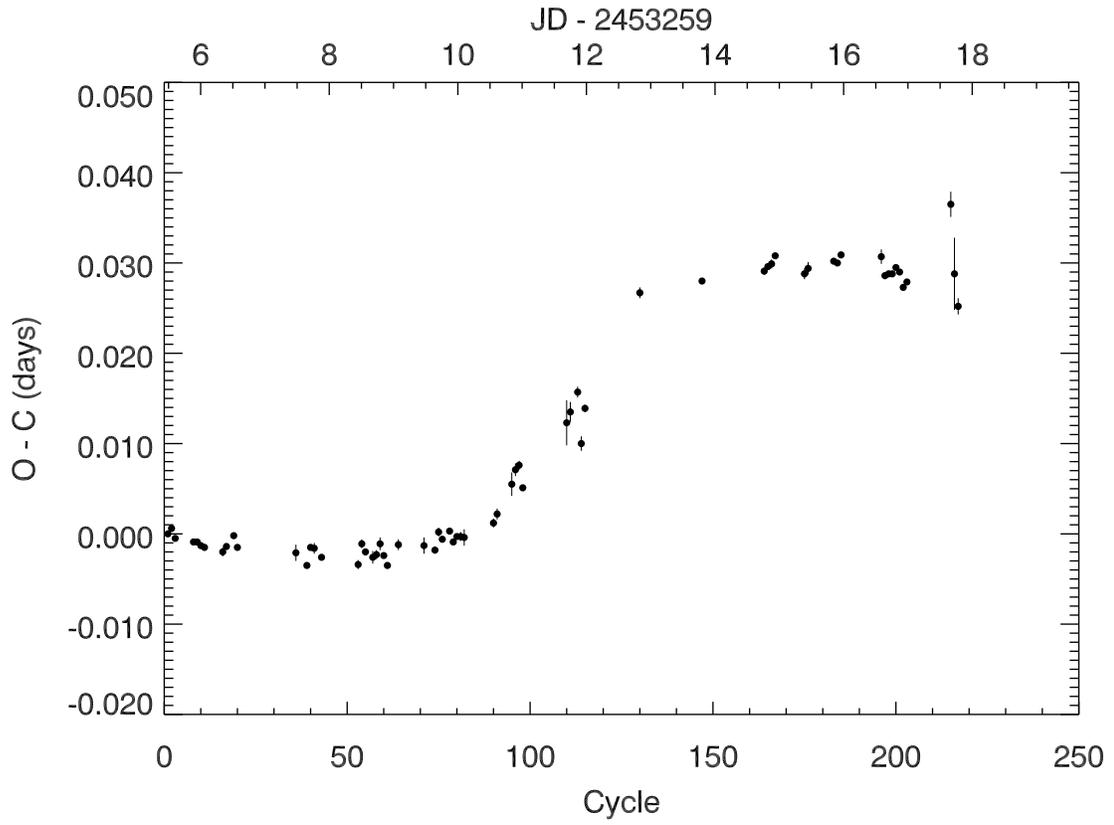}
\caption{$(O-C)$ diagram of the superhumps, assuming a period of 0.0569 days.
There is a significant increase in period beginning at JD 2,453,272 before the
superhumps return to the original period.}
\label{Fig04}
\end{center}
\end{figure}

We also used $(O-C)$ analysis to study the time-varying nature of the
superhumps.  The $(O-C)$ method has been used in the past to study superhump
evolution in SU UMa- and WZ Sge-type objects, and often shows substantial
changes over the course of individual outbursts \citep{howell96,pat02}.
Figure \ref{Fig04} shows an $(O-C)$ diagram of the outburst prior to the rapid
decline and first echo, using an ephemeris period of 0.0569 days.  Between 
JD 2,453,269 and 2,453,272 the phases of the superhumps steadily shifted 
until they reached a steady offset of approximately 0.03 days, or slightly more
than half a period.  This could indicate either a true change in period 
\citep{olech04}, or a change in superhump morphology.  This is a common feature
of dwarf novae that exhibit superhumps in outburst, and the $(O-C)$ diagram
is reminiscent of those of WZ Sge \citep{pat02} and AL Com \citep{howell96}.  
Visual inspection of Figure \ref{Fig02} shows that the superhumps are clearly 
changing in shape and amplitude as the outburst progresses.  

Unfortunately, we do not have any time-series observations for the earliest 
part of the 
outburst, and thus cannot make any statement on whether so-called ``orbital 
humps'' having the orbital period were present.  WZ Sge exhibited clear 
orbital humps during the first 12 days of the 
2001 outburst, as did AL Comae for the first several days.  Both stars were 
caught very close to the outburst onset, and time-series photometry commenced 
immediately.  There was a lag of at least five (and perhaps as much as ten) 
days between the onset
of the outburst of ASAS J002511+1217.2 and the commencement of time-series 
photometry.  An answer to this question will have to wait until the detection
of the next outburst of this system.

\subsection{Early decline}

\begin{figure}
\begin{center}
\includegraphics[width=0.95\textwidth]{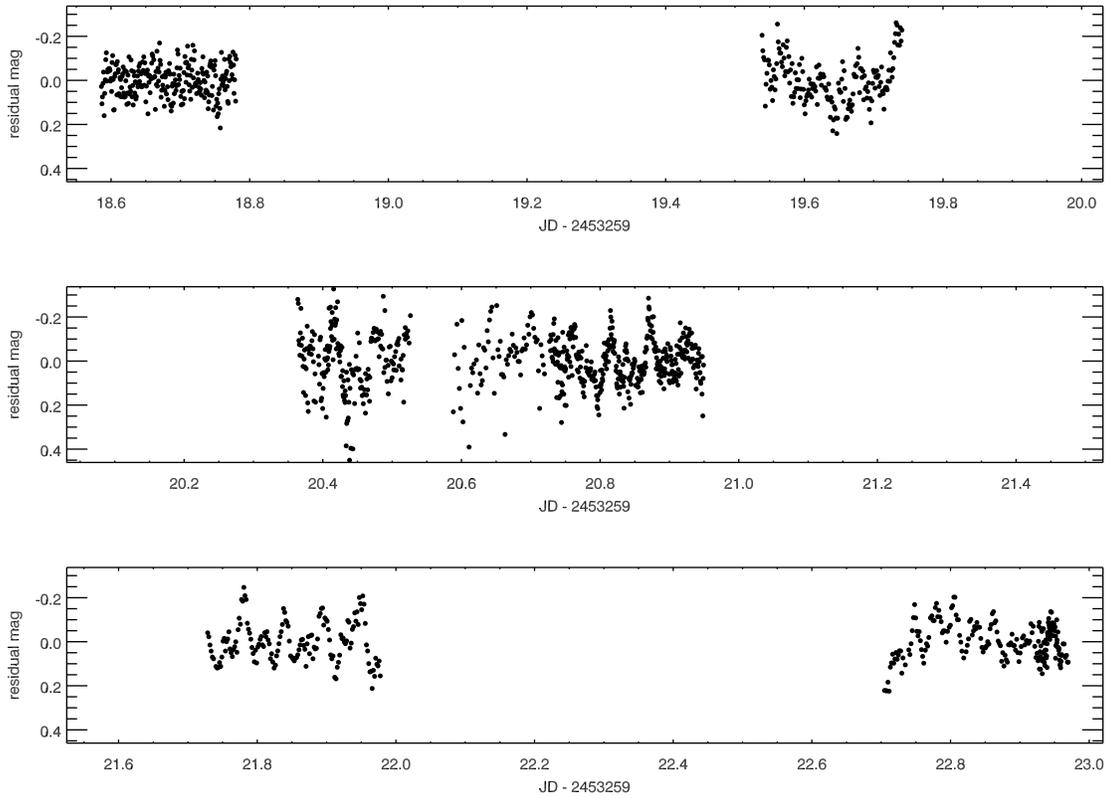}
\caption{Detrended light curve showing the early decline period prior to
the echo outburst.  The profiles of the humps are qualitatively different from
those of the superhumps observed during the superoutburst, and have a period of
$0.05666 \pm 0.00003$ days, below that of the late superhumps.  We do not 
believe these are orbital variations.}
\label{Fig05}
\end{center}
\end{figure}

\begin{figure}
\begin{center}
\includegraphics[width=0.95\textwidth]{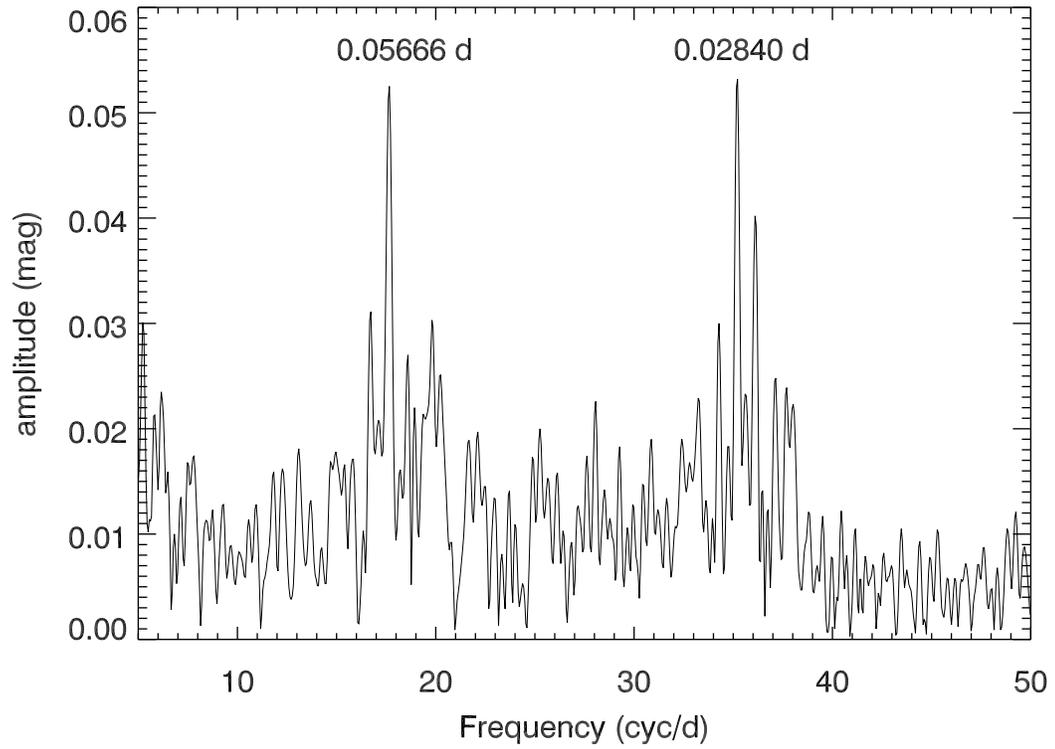}
\caption{Discrete Fourier transform of the data shown in Figure 5, showing the
period at 0.05666 days, along with its Fourier harmonic at half the period.
The double-peaked nature of the light variations is explained by the nearly
equal amplitudes of the two components.}
\label{Fig06}
\end{center}
\end{figure}

The system began to fade rapidly after JD 2,453,276.5, and spent nearly four 
days
at $V \sim 15$ prior to the single echo outburst.  The detrended time-series
measurements of ASAS J002511+1217.2 over this four day span are shown in Figure
\ref{Fig05}, and a Fourier spectrum of these data is shown in Figure
\ref{Fig06}.  Just prior to the end of the main superoutburst, the amplitude
of the superhumps dropped significantly, and they do not appear in the 
light curve during rapid fading.  If superhumps are caused by distortion of
the accretion disk \citep{pat02}, then the decreasing
amplitude may indicate the return of the disk to its normal shape prior to the
disk transitioning from its hot, outburst state to the cool state.

A new photometric period of $0.05666 \pm 0.00003$ days appears in the data 
during its time at $V \sim 15$, and the variations have a much smoother, 
double-peaked profile that is quantitatively different from that of the 
superhumps.  If this new period were the true orbital period, the resulting 
period excess $\epsilon = (P_{sh} - P_{orb})/P_{orb}$ would be 0.004, smaller 
than that of the shortest-excess WZ Sge star EG Cnc with $\epsilon = 0.007$ 
\citep{hellier01}.  We therefore suspect that this period is not the orbital 
period, but is another modulation related to superhumps, and that the as-yet 
undetected orbital period may be closer to the period minimum.

\subsection{Echo outburst}

One echo outburst occurred approximately five days after the end of the main
outburst, during which the star returned to $V \sim 12.8$ for less than one day 
before it rapidly declined to $V \sim 16$. Following this single echo, no 
other echoes were observed, and ASAS J002511+1217.2 began its slow decline to 
quiescence over the following 70 days.  Very weak variations are present
during the echo; a period of about 0.0568 days is marginally detected, close to
the period of the late superhumps.  \citet{golovin05} fit maxima to these
variations and integrated them into their $(O-C)$ analysis, indicating that
the resulting $(O-C)$ is similar to that of other WZ Sge-star outbursts like
WZ Sge itself \citep{pat02}.

The timing of the echo outburst is similar to that of EG Cnc, which underwent
a series of echo outbursts beginning several days after the initial decline
of the superoutburst.  However, ASAS
J002511+1217.2 exhibited only {\it one} echo.  There are several theories as
to the causes of echo outbursts, though all theories predict that the disk
must be heated over the thermal instability limit for a new outburst to
occur.  The occurrence of only one echo suggests that the accretion disk had
either been significantly drained of matter following the initial 
superoutburst or first echo, or had been cooled sufficiently that the reheating
mechanism could only increase the temperature beyond the instability limit one
time before the disk began a transition back to its quiescent state.  We
discuss these scenarios further in Section 4.

\subsection{Late outburst and return to quiescence}

The end of the first echo on JD 2,453,284 left ASAS J002511+1217.2 at 
$V \sim 16$, and the system never rebrightened significantly over the following
two months.  Time-series measurements were obtained through JD
2,453,303, with visual and CCD monitoring continuing through early 2005.  The
system declined by about one magnitude over the 70 days following the end of
the echo outburst, returning to quiescence at $V \sim 17$.  A few nights of
time-series CCD photometry were obtained by several observers following the
echo outburst.  We performed a Fourier analysis of the data, and found a signal
at 0.0570 days, nearly the same as the {\it superhump period}, and not the
supposed orbital period as one would expect.  This suggests that the disk
remained elliptical throughout outburst {\it and} quiescence.  Permanent 
superhumps are believed to be a feature of high mass-transfer systems, because 
the disk maintains a large enough radius to be tidally distorted and precessing
\citep{osaki96,rn00}.  However, the WZ Sge stars are believed to have a 
{\it low} mass-transfer rate.  \citet{tra05} recently found an SU UMa-type
system with what appeared to be negative superhumps at quiescence, possibly
due to a tilted accretion disk.  Photometry of ASAS J002511+1217.2 obtained
with a larger telescope during quiescence can determine whether superhumps 
are actually present, and whether there are orbital variations.

\section{Spectroscopy}

The spectroscopic data were acquired with the Apache Point Observatory
(APO) 3.5m telescope on the night of 2004 October 10 from 04:39:21 to
06:39:36 UT (JD 2,453,288.6940 to 2,453,288.7775).  A total of 15 exposures
were taken over this time interval, each with an exposure time of 360
seconds.  This allowed observations at an average separation of seven
minutes. The primary instrument used was the Double Imaging Spectrograph
(DIS) run in high resolution mode with a 1.5'' slit.
The DIS instrument is designed so that the incoming light is split by a
dichroic, and red and blue spectra are taken simultaneously on two
separate CCD cameras, allowing for an extended spectral range and improved
wavelength resolution.  The wavelength coverage was 3850 {\AA} to 5550 {\AA} 
with 0.6 {\AA} per pixel resolution on the blue images, and 5550 {\AA} to 7250
{\AA} with 0.8 {\AA} per pixel resolution on the red images.  The spectral
coverage provides adequate sampling of several strong emission lines.  

The raw spectra were processed using 
IRAF\footnote{Image Reduction and Analysis Facility, is
distributed by the National Optical Astronomy Observatories, which are
operated by the Association of Universities for Research in Astronomy,
Inc., under cooperative agreement with the National Science Foundation.}
subroutines to produce one-dimensional wavelength and flux calibrated
spectra in both wavelength regimes. The target exposure time of 6
minutes allowed for reasonable signal-to-noise ratios (S/Ns) such that our
orbital period uncertainties were not limited by the raw data resolution,
but by limits imposed by the total span of observations and the separation
interval of consecutive exposures.  A $V$ 
magnitude of 16.3 was estimated from the flux-calibrated spectra; this
agrees with the $V$-band time-series photometry obtained on JD 2,453,285 
and 2,453,290.
Coincident $R$-band time-series photometry obtained around JD 2,453,288.7 
has an average near 15.9.

\subsection{Spectroscopic results}

\begin{figure}
\begin{center}
\includegraphics[width=0.9\textwidth]{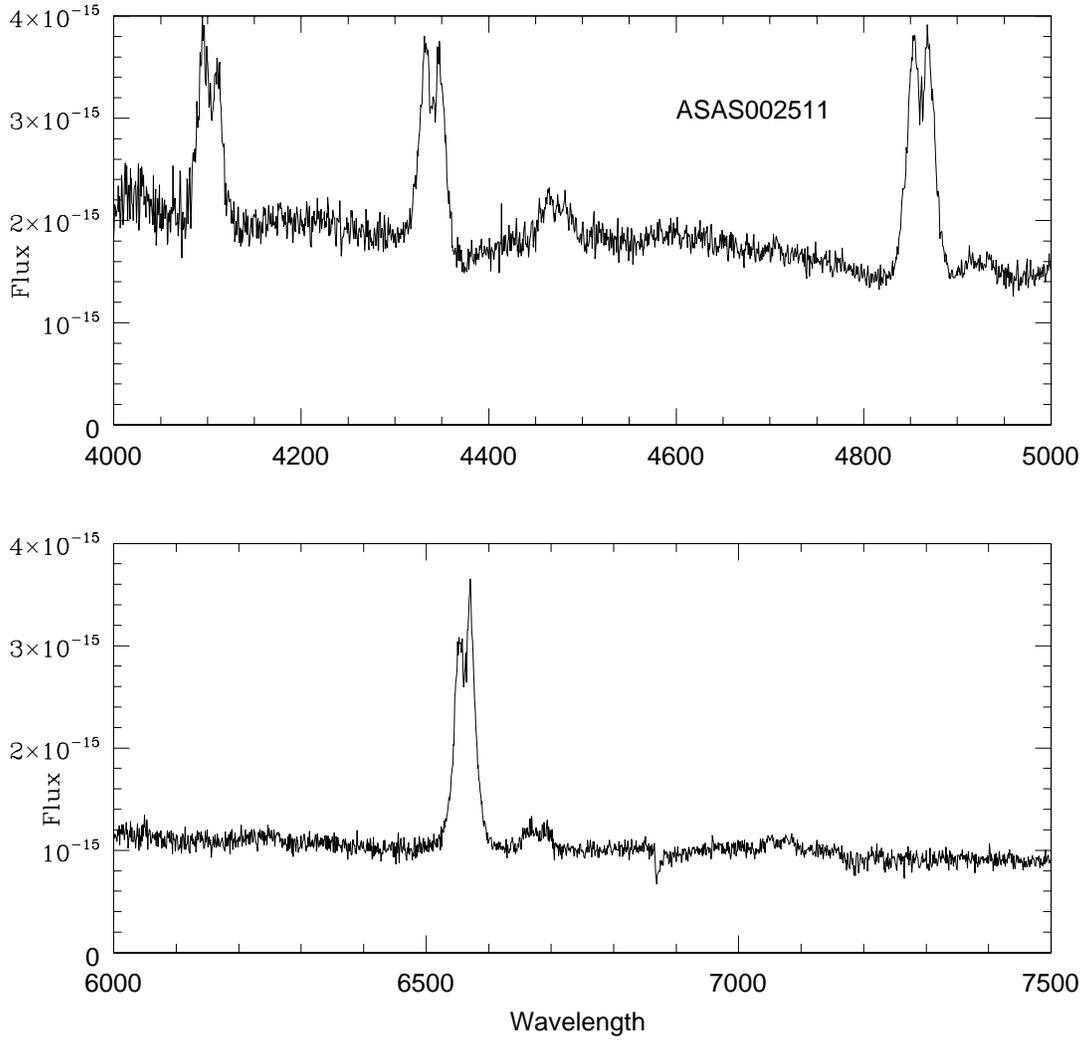}
\caption{Typical blue and red spectra, showing the broad,double-peaked lines.
The H$\alpha$ (6562.852 {\AA}), H$\beta$ (4861.342 {\AA}), and H$\gamma$ 
(4340.475 {\AA}) lines were all used for velocity measurements.}
\label{Fig07}
\end{center}
\end{figure}

Figure \ref{Fig07} shows sample red and blue spectra for ASAS 
J002511+1217.2.
All the spectra show prominent, broad Balmer lines in emission with a
central absorption that is typical for systems containing an accretion
disk. The average equivalent widths were 66, 39, and 24 {\AA} for
H$\alpha$, H$\beta$ and H$\gamma$ respectively. These values, together
with the widths of the line (full-width, zero-intensity measures of 75 {\AA} 
and 60 {\AA} for H$\alpha$ and H$\beta$ respectively) and the depth of the 
central absorption imply a fairly high inclination. 

\begin{figure}
\begin{center}
\includegraphics[width=0.9\textwidth]{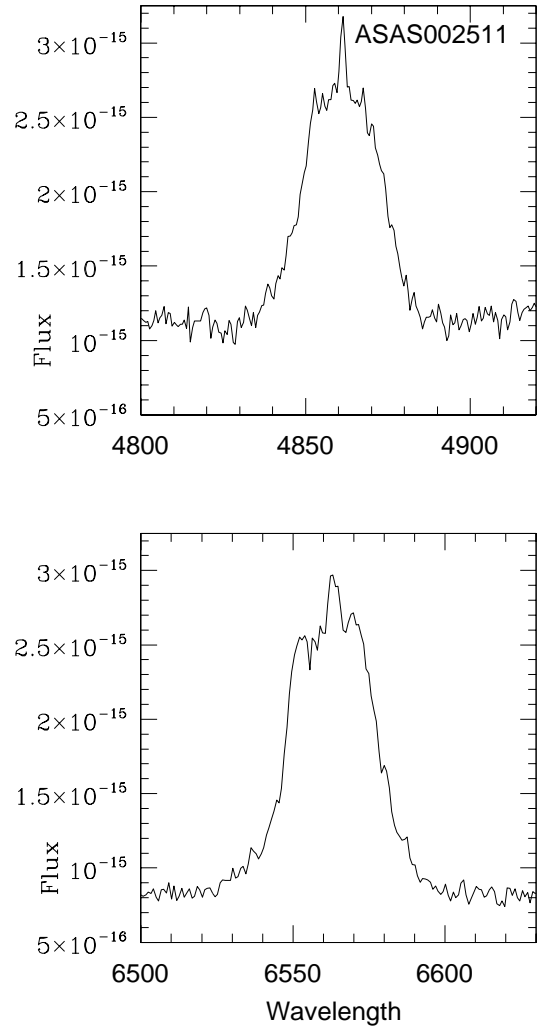}
\caption{An enlargment of the H$\alpha$ and H$\beta$ lines at phase
0.9, showing the narrow component structure that is likely due to
a hot spot on the outer edge of the accretion disk around the primary
star.}
\label{Fig08}
\end{center}
\end{figure}

Velocity measurements of the Balmer emission lines were made in order to
determine the orbital period of the system.  The H$\alpha$, H$\beta$, and 
H$\gamma$ lines all had sufficient signal to noise to allow for measurement of
their central wavelengths.  Initially, the IRAF centroid
fitting technique `e' was used in the $\textsc{splot}$ task to estimate
the line wavelengths in each spectrum over the course of the night.  These
wavelengths together with observation times were run through the IDL routine 
$\textsc{sinfit3}$ that performed a least-squares sinusoidal fit to
the data, and simultaneously output the calculated values and
uncertainties for period (P), semi-amplitude (K), phase offset $(\phi)$
and the velocity of the system $(\gamma)$. However, as one example shows
in Figure \ref{Fig08}, some of the lines had sub-structure that prevented a
consistent wavelength estimate.  Although the emission lines were strong
and statistically distinguishable from the continuum noise, narrow
components can distort the centroid calculation. 

\begin{figure}
\begin{center}
\includegraphics[width=0.9\textwidth]{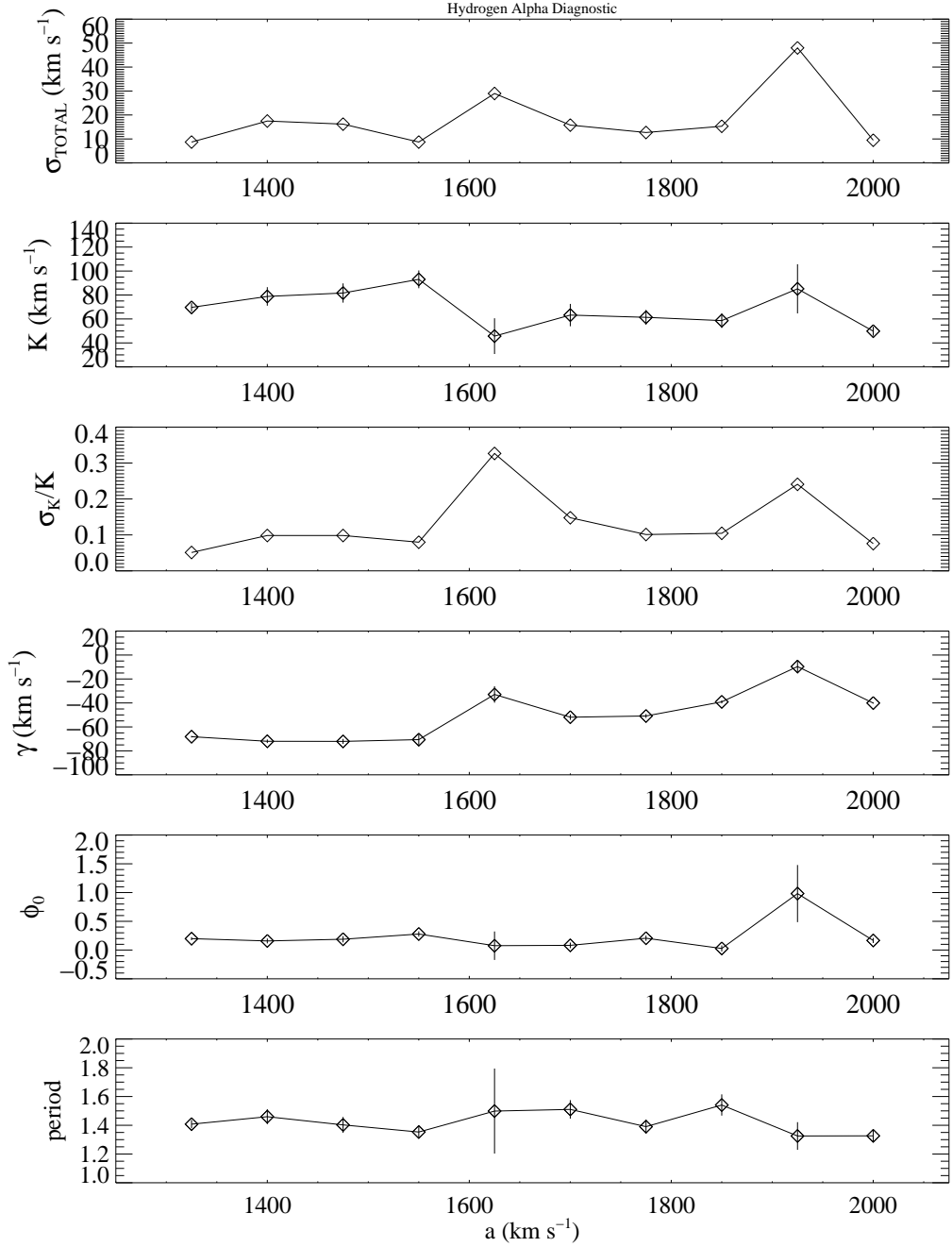}
\caption{A sample diagnostic plot from the double Gaussian output.  The 
velocities corresponding to a given double Gaussian separation are plotted 
with the characteristic parameters from the sinusoidal fit.  For this 
H$\alpha$ line, a minimum $\sigma_{total}$ corresponding to 
$a=1550\frac{km}{s}$ was chosen.}
\label{Fig09}
\end{center}
\end{figure}

A second velocity measurement was made of all spectra using a double
Gaussian routine that is optimized to measure the line wings \citep{shafter83}.
This method works best for broad lines with high signal to noise
ratios.  Shafter found that the optimum Gaussian separation width for
computing the central wavelength corresponds to an effective flux of about
$<f>=0.3$ or $\frac{1}{3}$ of the line flux.  Analyzing the velocities
closer to the wings of the emission lines rules out the possibility that
variations resulting from a hotspot located in the outer edge of the disk
could mislead a centroid finding program, as the wings correspond to the
faster rotating inner disk.  To allow maximum flexibility for a variety of
spectra, this method iterates through and produces a variety of Gaussian
separations as a function of velocity.  Each one of these Gaussian separations
is put into the IDL routine $\textsc{sinfit}$ which generates a radial velocity
curve where the errors can be minimized interactively.  By looking at a 
diagnostic plot (Figure \ref{Fig09}) of the sinusoidal fits for each 
of the radial velocity measurements, the fit with the lowest errors could be
selected for each of the emission lines.  The best estimate for the period
from this Double Gaussian method, could then be set as a fixed parameter
in our original IDL sinusoidal fitting routine $(\textsc{sinfit3})$ and
the error estimates could be compared with an unconstrained fit.
Additional analysis attempting to search for periodicity in the blue and
red continuum fluxes in the spectra showed no repetitive behavior. 

\begin{figure}
\begin{center}
\includegraphics[width=0.9\textwidth]{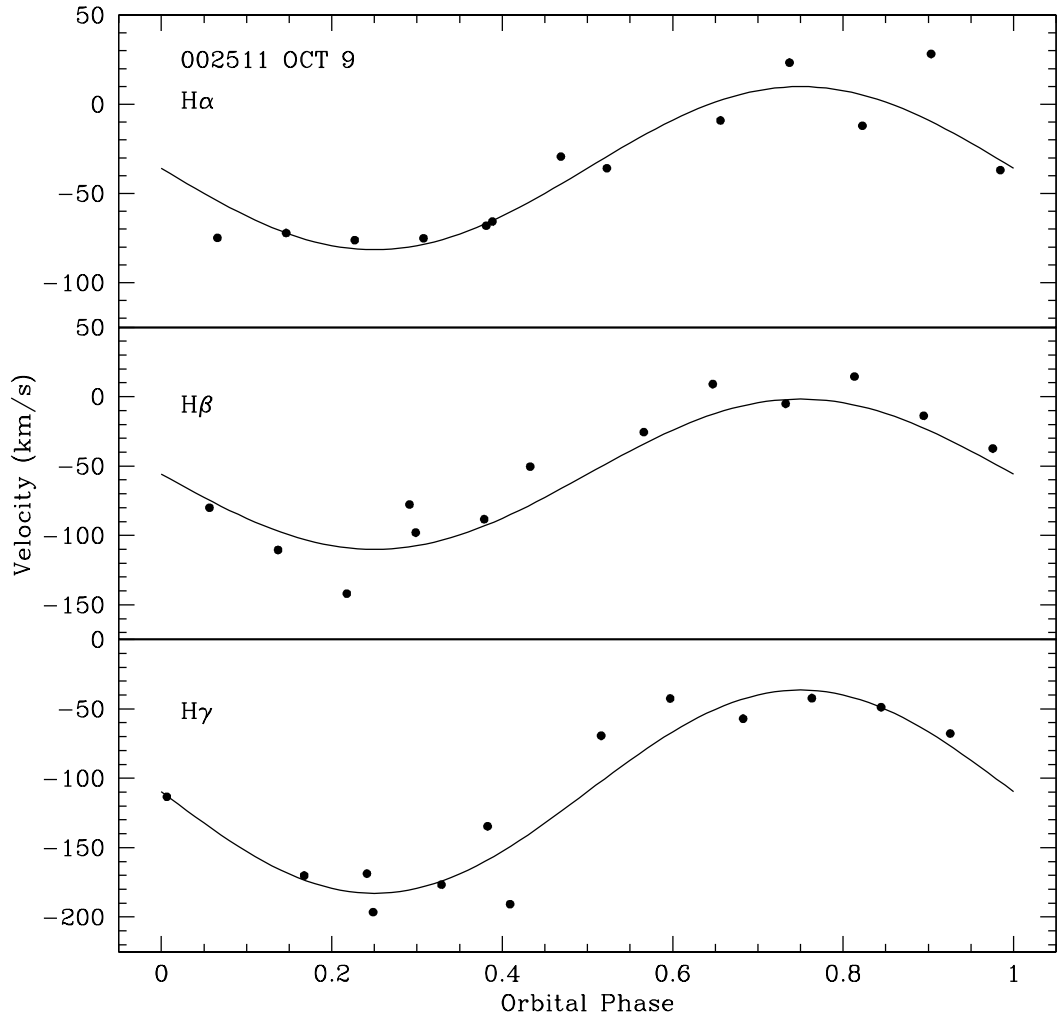}
\caption{The fits from the $\textsc{sinfit3}$ final run are shown over-plotted
on the velocity data points corresponding to the 
double Gaussian velocities with period fixed at 82 min.}
\label{Fig10}
\end{center}
\end{figure}

The final parameters from both of the analysis routines are presented for
comparison in Table \ref{tab01}. Since there was large disagreement in the 
periods
for the centroid analysis (which could be partly due to the differing
contributions of the narrow line component in the blue versus the red), we
fixed the period at the average of the Double Gaussian method for the 3
lines (82 min) and reran a final solution which is the bottom entry in
Table \ref{tab01}. The final fits to the data with this solution are shown 
in Figure \ref{Fig10}. 

\begin{figure}
\begin{center}
\includegraphics[width=0.9\textwidth]{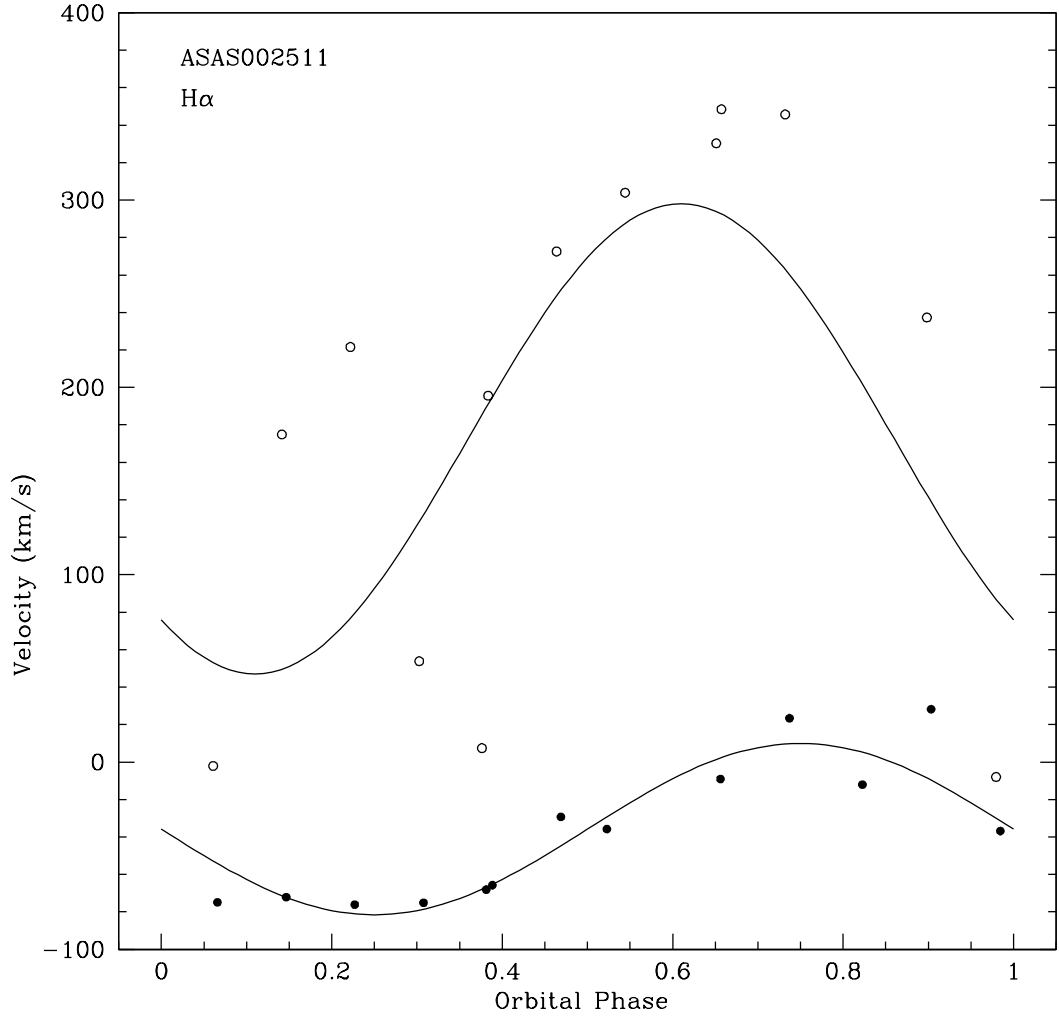}
\caption{Sinusoidal fit and data points for the narrow (open circles) and 
broad (filled circles) components of H$\alpha$.  Note the relative phase offset
of the two components and the amplitude increase of the narrow component over 
that of the broad.}

\label{Fig11}
\end{center}
\end{figure}

Analysis of the narrow components of the Balmer emission features via the
same IRAF centroid routine was attempted to estimate the location of a
hotspot. The phase offset between the narrow and broad components allows us to
quantitatively estimate where the mass stream impacts the accretion disk. 
However, measurement of the narrow component
is difficult as it could not be accurately located at all phases. The
result of our measurements of the velocities for the narrow component in
H$\alpha$ is shown overplotted with the broad component in Figure \ref{Fig11}. 

Although an exact origin of the narrow component cannot be calculated, the
narrow component is most visible around an orbital phase of 0.9. This
rules out irradiation of the secondary star as an origin for this
component, because at an orbital phase of 0.9 the secondary is near inferior 
conjunction and we are viewing the
side of the secondary that is away from the white dwarf. This, together
with the phase offset and large velocity amplitude of the narrow component
(Figure \ref{Fig11}) argues for an origin in a hot spot due to the impact of the
accretion stream on the outer edge of the disk. 

\section{ASAS J002511+1217.2: A new WZ Sge system}

We argue that the observational evidence gathered so far indicates that
ASAS J002511+1217.2 is a WZ Sge star.  The outburst amplitude of seven
magnitudes (based upon the initial ASAS detection at $V \sim 10.4$ and the 
Hamburg Quasar Survey pre-outburst detection at $V \sim 17.3$) is marginally
larger than the maximum of about six magnitudes seen in SU UMa-type dwarf 
novae.  This star is also a 2MASS point source, and the $(J-H)$ and $(H-K)$ 
colors of $0.729 \pm 0.239$ and $0.357 \pm 0.292$ respectively suggest a 
late-K, early-M companion (see Figure 4 in \citet{hoard02}).
The period as derived from superhumps and late-outburst photometry
is short, and similar to other known WZ Sge
systems, although SU UMa stars can have similarly short periods.  The
superhump evolution is nearly identical to that observed in other WZ Sge
stars.  The overall outburst lightcurve is similar to those of WZ Sge
systems, including the presence of an echo outburst, and a very long
decay time to true quiescence ({\it at least} 90 days from peak to 
$V \sim 17$).  The echo outburst is particularly striking evidence, as
echoes have not been detected in SU UMa stars; only in WZ Sge stars.  The
single echo detected in ASAS J002511+1217.2 was separated by a few days from
the rapid decline of the main outburst, as is seen in the most prominent
echo outburster, EG Cnc.

However, further information is needed since we lack a clear understanding of 
the development of the superhumps, and haven't determined the orbital period.  
The period of $0.05666 \pm 0.00005$ days observed during the short 
quiescent interval between the main outburst and the first echo is accurate, 
and does not appear to be a sidelobe misidentification.  The 0.0570-day period
present after JD 2,453,288 matches the supposed superhump period observed during
the main outburst.  It is possible that the early variations observed during 
the main outburst were {\it orbital humps}, before the disk had begun to
precess, but such humps typically appear very early in outburst. Our 
time-series photometry does not begin until {\it at least} five days after
the star reached maximum, and it is possible that the true maximum occurred 
as much as five days before the ASAS detection because the last non-detection
was made on JD 2,453,254.

The presence of only one echo outburst may indicate a low mass transfer rate.  
Echo outbursts may be due to several things, including a temporary increase in
mass transfer caused by heating of the secondary \citep{pat98,bmh02},
or temporary reheating episodes of a still-elliptical, post-superoutburst 
disk while it drains \citep{sg01,ommh01,matthews05}.  The presence of superhumps
during echo outbursts that are in phase with those of the main outburst 
\citep{golovin05} suggests that the main outburst ends while the disk is 
still elliptical.  If echo outbursts are triggered by mass draining from the
outer disk, then the single echo implies that this reservoir was sufficiently
drained by the original outburst and first echo to preclude any additional
echoes.

\section{Conclusions}

Our photometric and spectroscopic study of ASAS J002511+1217.2 has revealed
a system showing all the properties of a low accretion rate dwarf nova.  The
bulk of the evidence (large outburst amplitude, an echo outburst, superhumps,
long decline to quiescence, and a suspected short orbital period) justifies
a WZ Sge classification rather than an SU UMa system.  It does not appear that
this system has eclipses, although the breadth and deep doubling of the 
emission lines implies a fairly high inclination.  Longer time-resolved 
spectroscopy can be used to pin down the orbital period to higher precision.
This can then be compared with the superhump period to obtain a period excess 
and hence a mass ratio and estimate of the mass of the secondary, as has been 
done for other WZ Sge systems \citep{pat98}.  Further time-resolved photometry
during quiescence can determine if the continuum is modulated at the orbital
period due to structures in the disk.  Continued monitoring of this system,
both by the amateur community and by automated systems such as ASAS
\citep{poj02} can catch future outbursts and determine if the single
echo outburst is typical of this system, and related to its particular mass
transfer rate.  This system is bright enough at maximum to be well-within the
range of both visual and CCD observers.

\acknowledgements

We acknowledge with thanks the variable star observations from the AAVSO
International Database contributed by observers worldwide and used in this
research. PS acknowldeges support from NSF grant AST 02-05875.  The AAVSO
thanks the Curry Foundation for their continued support of the AAVSO 
International High-Energy Network.  This publication makes use of data products
from the Two Micron All Sky Survey, which is a joint project of the University
of Massachusetts and the Infrared Processing and Analysis Center/California 
Institute of Technology, funded by the National Aeronautics and Space 
Administration and the National Science Foundation.

\begin{deluxetable}{cccccccc}
\tablewidth{0pt}
\tablecaption{\textsc{Radial Velocity Solutions}}
\tablehead{
\colhead{Line Measurement} & \colhead{Line} &
\colhead{P} & \colhead{$\sigma_{p}$/P} & \colhead{K} &
\colhead{$\gamma$} & \colhead{$\phi_{0}$} & \colhead{$\sigma_{total}$} \\
\colhead{Algorithm} & \colhead{} & \colhead{(min)} & 
\colhead{(km s$^{-1}$)} & \colhead{(km s$^{-1}$)} & 
\colhead{} & \colhead{} } 
\startdata
Iraf Centroid & H$\alpha$ & 80.4 & - & 46 $\pm$ 5 & -36 $\pm$ 1 & 0.17 & 14 \\
 & H$\beta$ & 95.1 & - & 61 $\pm$ 5 & -55 $\pm$ 1 & -0.4 & 14 \\
              & H$\gamma$ & 91.1 & - & 78 $\pm$ 5 & -109 $\pm$ 1 & -0.2 & 14 \\\hline
Double Gaussian & H$\alpha$ & 81.2 & 0.03 & 93 $\pm$ 7 & -71 $\pm$ 5 & 0.28 & 9 \\
 & H$\beta$ & 83.9 & 0.09 & 41 $\pm$ 9 & -48 $\pm$ 2 & 0.14 & 23 \\
                & H$\gamma$ & 80.2 & 0.08 & 50 $\pm$ 2 & -95 $\pm$ 3 & 0.11 & 24 \\\hline
Double Gaussian & H$\alpha$ & 82 & - & 46 $\pm$ 5 & -36 $\pm$ 1 & 0.1 & 14 \\
 & H$\beta$ & 82 & - & 54 $\pm$ 7 & -56 $\pm$ 1 & 0.2 & 19 \\
Period fixed at 82 & H$\gamma$ & 82 & - & 73 $\pm$ 7 & -110 $\pm$ 2 & 0.2 & 20 \\
\enddata
\tablecomments{Comparison of the radial velocity data using the two
measurement algorithms discussed in the text.  The final values
should be taken from the last rows in the table.}
\label{tab01}
\end{deluxetable}

\end{document}